\documentstyle[12pt]{article}

\begin{document}
\def\bq{\begin{equation}}
\def\eq{\end{equation}}
\begin{flushright}
{June, 1998}\\
UT-Komaba-98/14
\end{flushright}

\begin{center}
{\large\bf
{DENSITY OF STATE IN A COMPLEX RANDOM MATRIX THEORY WITH 
EXTERNAL SOURCE }} \end{center}

\begin{center}
{\bf{  S. Hikami and R. Pnini}} \end{center}
\vskip 2mm
\begin{center}
 Department of Pure and Applied Sciences, University of Tokyo\\
Meguro-ku, Komaba, Tokyo 153, Japan\\ 
\end{center} \vskip 3mm
\begin{abstract}
  The density of state for a complex $N\times N$
  random matrix coupled to an external
  deterministic source is considered for a finite N, 
  and a compact expression in an integral representation
  is obtained.

\end{abstract}
\newpage

 The random matrix theory, in which  the eigenvalues of random matrix are 
 complex, may find some
 applications. For example,
   the two-dimensional electron systems under a strong
 magnetic field [1] or  the study of neural network [2] is similar 
 to the random matrix theory. 
  
  In a long time ago, Ginibre [3] considered  the complex random matrix 
  theory and he
 obtained a density of state $\rho(z)$ ( $z = x + i y$ );
 inside the circle in  complex plane, the density of states
  $\rho(z)$ becomes uniformly
 flat and is vanishing outside the circle. This is a generalization
 of Wigner's semi-circle law in the large N limit for the complex case.

 Recently, the random matrix theory with an external source
  has been investigated [4-16]. The external source is a deterministic, 
  and non-random matrix, coupled to a random matrix.
   It has been discussed for a Hermitian random 
 matrix [4-8] and for a chiral case [9]. Feinberg and Zee has studied
 the complex random matrix with an external source in the large N limit
 [10]. The asymmetric random matrix with external source
 has been  also studied [10,11]. 
 In the large N limit, the boundary of the density of state in the complex 
 plane
 may be obtained by several methods.
 However, the expression for the density of state, in a finite N,
 is much harder for the external source problem.
 It is known that 
 there appear interesting transitions of opening a gap 
 by tuning the external source
 [5,13]. It may be crucial to obtain an exact expression 
 for the density of state in a finite N for such problems.
 
 In this letter, we study a 
 complex random matrix which couples to an  external source matrix. We 
 generalize the previous works for the real eigenvalues [6,9] to this 
 complex eigenvalue case.
 The density of state $\rho(z)$ for the complex eigenvalues $z$ is given by 
 \bq\label{1.1}
 \rho(z) = {1\over{N}} < \sum_{i=1}^N 
 {\rm tr} \delta ( x - {\rm Re} \lambda_i )\delta ( y - {\rm Im}
 \lambda_i) >
 \eq
 where $\lambda_i$ is an eigenvalue of a complex matrix $M$,
  which couples to the external
 source matrix $A$ through the following probability distribution for the 
 present case,
 \bq\label{1.4}
 P_A(M) = {1\over{Z_A}} e^{- N {\rm tr} M^{\dag} M + N {\rm tr}
 (M^{\dag} A + A^{\dag} M)}
 \eq

 It has obtained by Ginibre [3] for a finite N, and $A = 0$ as
 \bq\label{1.3}
 \rho(z) = {1\over{\pi}} \sum_{n=0}^{N-1} {N^n|z|^{2 n}\over{n!}} e^{-N |z|^2}
 \eq
 where we write the result in which  the radius of the disk
 is unity in the 
 large N limit as a normalization.
 To evaluate the density of state (\ref{1.1}), it is useful to consider
 a chiral Hamiltonian,
 \bq\label{1.5}
 H = \left(\matrix {0 & M^{\dag} \cr M & 0\cr}\right) + 
 \left(\matrix {0 & A^{\dag}\cr
 A & 0\cr}\right)
 \eq
 where $M$ is a complex matrix. 
 We denote the density of state of this chiral Hamiltonian by
 $\rho_{ch}(\lambda)$, 
 \bq\label{1.6}
 \rho_{ch}(\lambda) = {1\over{2 N}} < {\rm tr} \delta (\lambda - H) >
 \eq
 where the probability distribution is $P(H) = {1\over{Z}}
 \exp[ - N {\rm tr} M^{\dag} M ]$.
 Note that the eigenvalues of $H$ are always real and appear in pairs
of  positive and negative values. Due to this chirality
 of the eigenvalues, the density of state $\rho_{ch}(\lambda)$ is 
 equal to
 \bq\label{1.7}
 \rho_{ch}(\lambda) = |\lambda| \tilde \rho(\lambda^2)
 \eq
 where
 \bq\label{1.8}
 \tilde \rho(r) = {1\over{N}} < {\rm tr} \delta (r - M^{\dag} M ) >
 \eq
 in which the average distribution probability $P(M)$ is same as $P_A(M)$ in 
 (\ref{1.4}).
 
 As noticed by Feinberg and Zee [10], the density of state $\rho(z)$ 
 in (\ref{1.1}) is obtained from the expression of the density of state
 $\rho_{ch}(\lambda)$ by the shift of $A$ in (\ref{1.4}) as $A \rightarrow
 A - z I$.
 Using the well-known expressions for the complex delta-function $\delta
 (z) = \delta (x) \delta (y)$, z = x + i y,
 \bq\label{1.9}
 \delta (z - z_0) = {1\over{\pi}} {\partial\over{\partial z^{*}}}({1\over{
 z - z_0}})
 \eq
 \bq\label{1.9a}
 \pi \delta(z) = \partial_z\partial_{z^{*}} {\rm log}(z z^{*}),
 \eq
 we have 
 \bq
 \rho(z) = {1\over{\pi}} \partial_z\partial_{z^{*}}< {1\over{N}} {\rm tr}
 {\rm log} (z - M)(z^{*} - M^{\dag})>
 \eq
 Using a dispersion relation between Green function and the 
 density of state $\rho_{ch}(\lambda)$, we get
 \begin{eqnarray}\label{1.10}
 \rho(z) &=& - {2 i\over{\pi}}
 \int_{0}^{\infty} ds \partial_z \partial_{z^{*}}( \int_{-\infty}^{\infty}
 {\rho_{ch}(\lambda)\over{is - \lambda}}d\lambda )
 \nonumber\\
 &=& {4\over{\pi}} \partial_z\partial_{z^{*}} \int_0^{\infty} ds
 \int_{0}^{\infty} d\lambda {\lambda s\over{\lambda^2 + s^2}}
 \tilde \rho(\lambda^2)
 \end{eqnarray}
 in which the external source $A$ is shifted as
 $A = diag ( |a_1 - z| e^{i \theta_1}, ..., |a_N - z| e^{i \theta_N} )$.
 In the large N limit, this $\tilde \rho(\lambda^2)$ was obtained by
 a diagrammatic analysis, and the density of state $\rho(z)$ was
 obtained by this procedure [10]. We consider here the finite N case, 
 not in the large N limit, by calculating the chiral $\rho_{ch}(\lambda)$
 with an external source through Itzykson-Zuber integral [17].
 
 A complex matrix $M$ is decomposed as
 \bq\label{1.11}
 M = U X V
 \eq
 where $U$ and $V$ are  unitary matrices and $X$ ia a diagonal
 matrix. Since the number of real variables is $2 N^2$ for $M$, 
 $N^2$ for $U$,$V$ and $2 N$ for X, we have $2 N$ conditions on 
 $U$,$V$ and $X$.
 It is possible to put the condition that
 the diagonal element of $V$ is real, and 
 $X = diag (x_1, ... , x_N)$, $x_i$ is real, $x_i \ge 0$. Note that 
 $x_i$ is not an eigenvalue of $M$, but $x_i^2$ is an eigenvalue of
 $M^{\dag}M$. $x_i$ is called a singular value of $M$.
 
 Itzykson-Zuber integral for this case is known [18-20],
 \bq\label{1.12}
 \int dU dV e^{{\rm Re} ( {\rm tr} U X V Y )} =
 {(2\pi)^{N^2}\over{N!}}
{{\rm det} [ I_0(x_i y_j)]\over{\Delta (x^2) \Delta (y^2)}}
 \eq
 where $Y = diag( y_1, ... , y_N)$ and $\Delta (x^2) = \prod_{i<j}
 ( x_i^2 - x_j^2)$,  which is a Van der Monde determinant.
 This Itzykson-Zuber integral is obtained by 
 applying a Laplacian of $M$ to (\ref{1.12}). This Laplacian 
 reduces to the diagonal one, and (\ref{1.12}) is a zonal spherical
 function.
 
 The external source $A = diag ( |a_1 - z| e^{i \theta_1}, ... ,
 |a_N - z| e^{i \theta_N} )$ is decomposed as $A =  Y \tilde U$, 
 $\tilde U = diag ( e^{i \theta_1}, ... , e^{i \theta_N})$ and 
 $Y = diag( |a_1 - z|, ... , |a_N - z|)$.
 This phase unitary matrix $\tilde U$ can be  absorbed in $U$.
 Thus, we have a diagonal matrix element $y_i = |a_i - z|$ in $Y$.
 
  Using the contour representation method by Kazakov [21], 
  and taking the same procedure
by Br\'ezin, Hikami and Zee (BHZ)[9], we evaluate the evolution
operator $U_A(t)$, which is a Fourier transform of the density 
of state $\tilde \rho(\lambda)$ in (\ref{1.7}), by 
noting that $x_i^2$ is an eigenvalue of $M^{\dag}M$,
\begin{eqnarray}\label{1.13}
   \tilde \rho(\lambda) &=& \int_{-\infty}^{\infty}
{dt\over{2 \pi}} e^{- i t \lambda} U_A(t)
\\
U_A(t) &=& {1\over{N}}< {\rm tr} e^{i t M^{\dag} M} >
\nonumber\\
&=& {1\over{N Z_A}}\sum_{\alpha}^{N}\int_0^{\infty}
dx \prod_{i=1}^N x_i {\Delta(x^2)\over{\Delta(y^2)}}
{\rm det} [ I_0(2 N x_i y_j) ] e^{- N \sum x_i^2 + i t x_{\alpha}}
\end{eqnarray}
The coefficient of (\ref{1.13}) is not important since
we normalize $U_A(t)$ as $U_A(0) = 1$.
This expression is similar to the previous results [6,8]
 except that we have
a modified Bessel function $I_0(2 N x_i y_j)$ instead of $e^{ N x_i y_j}$
as an element of the determinant. The modified Bessel function
has an integral representation as
\bq\label{1.14}
I_0(2 \sqrt{a}) = \int_{-\pi}^{\pi} {d \theta\over{2 \pi}}
e^{e^{i\theta} + a e^{-i\theta}}
\eq

Keeping the notation $y_i^2 = |a_i - z|^2$, we find the integral
representation for $U_A(t)$ as
\begin{eqnarray}\label{1.16}
U_A(t) &=& {1\over{N}}\int_{-\pi}^{\pi}{d\theta\over{2 \pi}}
\int_0^{\infty} dq \oint{du\over{2\pi i}}
{1\over{(1 - { i t\over{N}} - u)(f - ue^{i\theta})}}
 e^{- q + i \theta + (1 - u)
e^{i \theta}}
\nonumber\\
&\times& 
\prod_{i=1}^{N} ({f - N y_i^2\over{u e^{i \theta} - N y_i^2}})
\end{eqnarray}
where $f = ({i t   + N u\over{N - i t - N u}}) q$. 
 The contour integral over $u$
is reduced to evaluation of the residue at
 the pole $u = N y_i^2 e^{- i \theta}$.
The integration over $q$ is intoduced for the absorption of a combinatorial
factor $k!$, which appears in the $x_i$ integral in (\ref{1.13}). 

One can easily find that when there is no external source $y_i = 0$,
the expression $U_{A = 0}(t)$ in (\ref{1.16})
 reduces to the result of BHZ [9].
The density of state $\tilde \rho(\lambda)$ is given 
by the shift of $t \rightarrow N (t + i u)$,
\bq\label{1.18}
\tilde \rho(\lambda) = \int_{-\infty}^{\infty} {dt\over{2 \pi}}\oint{du\over{2 
\pi i}}
\oint{dv\over{2 \pi i}}\int_0^{\infty}dq 
{e^{- q - i N t \lambda + N u \lambda - v (1 - {1\over{u}})}
\over{u ( 1 - i t) ( f - v)}} \prod_i ({f - N y_i^2\over{v - N y_i^2}})
\eq
where $f = i t q/(1 - i t)$, and  we have made a change of variable 
$e^{i \theta} = v/u$.
Now we consider the density of state $\rho(z)$ for the complex
matrix through (\ref{1.10}).  We 
replace a factor as
\bq\label{1.19a}
    {1\over{s^2 + \lambda^2}} = \int_0^{\infty} e^{- \alpha ( s^2 +
\lambda^2) } d\alpha
\eq
Inserting (\ref{1.18}) into (\ref{1.10}), 
and 
 doing the integration over $s, \lambda$ and $t$, we obtain
  by the change of variables, $u \rightarrow u + \alpha/N$,
  $q \rightarrow (1 - u) q/u$ and $\alpha \rightarrow N \alpha u$
  and $\beta = \alpha/(\alpha + 1)$,
 \bq\label{1.19}
\rho(z) = - {1\over{\pi N}}\partial_z \partial _{z^{*}}
[ \int_0^{1} d\beta \oint {du\over{2\pi i}}\oint
{dv\over{2 \pi i}}\int_0^{\infty} dq {
e^{- {1 - u\over{u}}q - v + {v\over{u}}(1 - \beta)}
\over{\beta u^2 ( 
q - v)}}\prod_{i = 1}^N ({q - N y_i^2\over{v - N y_i^2}})]
\eq
where the contour of the integration of $v$ is around $N y_i^2$ and
the contour of $u$ is around $u = 1$, which appears as a pole
after the integration of $q$.

If $f(q)$ is a polynomial of $q$, we are able to prove that
\bq
 \oint{du\over{2 \pi i}}\int_0^{\infty} dq
  {1\over{u^2}}e^{{v(1 - \beta)\over{u}}}
  e^{-{(1 - u)q\over{u}}} f(q)
 = - e^{v(1 - \beta)} f( v(1 - \beta))
\eq
 Thus, the integrations over $q$ and $u$ can be done, and 
 we finally obtain by the shift $v \rightarrow N v$,
 \bq\label{1.21}
  \rho(z) = {1\over{\pi N}}\partial_z \partial _{z^{*}}
  [ \int_0^{1} d\beta \oint{dv\over{2\pi i}} {e^{-\beta N v}\over{\beta^2 v}}
  \prod_{i=1}^N ( 1 - {\beta v\over{v -  y_i^2}} ) ]
 \eq
 where  contours are taken around all $y_i^2$.  

It is easy to write down the explicit form for the small N.
We have
\begin{eqnarray}
  \rho(z) &=& {1\over{\pi}} e^{- |a_1 - z|^2}  \hskip10mm
( N = 1)\nonumber\\
&=& {1\over{\pi}} [ e^{-2 |a_1 - z|^2} + e^{- 2 |a_2 - z|^2}
- {1\over{4}}\partial_z\partial_{z^{*}}( {e^{- 2|a_1 - z|^2} - e^{- 2|a_2 - 
z|^2}
\over{|a_1 - z|^2 - |a_2 - z|^2}})] \hskip5mm(N = 2)
\nonumber\\
&=& {1\over{\pi}}[ \sum_{i=1}^3 e^{- 3 y_i^2} - {1\over{3}}
\partial_z\partial_{z^{*}}
\oint {du\over{2 \pi i}} {e^{-u}( 1 + 3 \sum_{i=1}^3 y_i^2 - 2 u)
\over{\prod_{i=1}^3 ( u - 3 y_i^2)}}] \hskip5mm (N = 3)
\end{eqnarray}
where $y_i^2 = (a_i - z)(a_i^{*} - z^{*})$.
It is also easy to see, 
when we put $a_i = 0$, we obtain $y^2 = z^{*}z$,  
and by the differentiation for $z$ and $z^{*}$,
(\ref{1.21}) becomes
\bq\label{1.30}
\rho_N(z) = {1\over{\pi}} \int_0^1 d\beta \oint {dv\over{2 \pi i}}
( N y^2 v - {1\over{\beta}}) ( 1 - {\beta v\over{v - 1}})^N
e^{-\beta N v y^2}
\eq

If we write $N y^2$ by $s^2$, and put $I_N(s) = \rho_N(z) - \rho_{N - 1}(z)$,
we find $I_N(s) = {s^{2(N-1)}\over{(N - 1)!}} e^{- s^2} /\pi$.
This agrees with Ginibre's result (\ref{1.3}). It is immediate
to obtain the boundary of the density of state from (\ref{1.21}) by
the saddle point equation. Taking the derivative of the
exponent in the large N limit by $\beta$, and putting $\beta = 1$, which
is an end point of the integral, we have as a boundary
curve $x^2 + y^2 = 1$ for Ginibre
case, and $x^4 + y^4 + 2 x^2 y^2 - 3 x^2 + y^2 = 0$ for the case when
the external source eigenvalues are $a_i = \pm 1$, $N/2$ times degenerated.

\vskip5mm
We are grateful to  Professors Edouard Br\'ezin and Anthony Zee for 
the useful discussions at 1st CREST workshop (Tokyo, September1997). This 
work was supported by the CREST of JST. S.H. acknowledges  a 
Grant-in-Aid for Scientific Research by the Ministry of Education, Science
and Culture.

\end{document}